\documentstyle[12pt,aasms4]{article}

\newcommand{\be}{\begin{equation}}
\newcommand{\ee}{\end{equation}}

\begin{document}
\title{Short-Term Dynamical Interactions Among Extrasolar Planets} 

\bigskip
\author{Gregory Laughlin$^1$, John E. Chambers$^{1,2}$} 

\bigskip
\affil{$^1$ NASA Ames Research Center, Moffett Field, CA 94035}

\affil{$^2$ Armagh Observatory, Armagh, BT61 9DG, UK}

\begin{abstract} 

We show that short-term perturbations among massive planets in
multiple planet systems can result in radial velocity variations of
the central star which differ substantially from velocity variations
derived assuming the planets are executing independent Keplerian
motions.  We discuss two fitting methods which can lead to
an improved dynamical description of multiple planet systems.  In the
first method, the osculating orbital elements are determined via a
Levenberg-Marquardt minimization scheme driving an N-body
integrator. The second method is an improved analytic model in which
orbital elements are allowed to vary according to a simple model for
resonant interactions between the planets.  Both of these methods can
determine the true masses for the planets by eliminating the $\sin i$
degeneracy inherent in fits that assume independent Keplerian motions.
We apply our fitting methods to the GJ 876 radial velocity data (Marcy
et al. 2001), and argue that the mass factors for the two planets are
likely in the 1.25-2.0 range.

\end{abstract}


\section{Introduction} 

About two thousand nearby stars are now being surveyed for
periodic radial velocity variations which indicate the presence of
extrasolar planets (Marcy, Cochran \& Mayor 2000). These search programs
have been highly successful, and to date they have discovered almost 
sixty extrasolar planets.

Recently, systems with more than one planet have been found, and four
($\upsilon$ Andromedae, GJ 876, HD 83443, and HD 168443) are now
known.  GJ 876 (Marcy et al 2001) provides an especially interesting
case.  In this system, a combined, two-Keplerian fit to the radial
velocity data (see Table 1), suggests that the star is accompanied by
two planets on orbits having a nearly commensurate 2:1 period ratio.
The amplitudes of the star's radial velocity variations suggest
minimum masses of 0.56 $M_{\rm JUP}$ for the inner planet, and 1.89
$M_{\rm JUP}$ for the outer planet. GJ 876, an M dwarf star with an
estimated mass of $0.32 \pm 0.05 M_{\odot}$ (Marcy et al 2001), is the
lowest mass star known to harbor planets.

For these orbital parameters, the mutual perturbations of the two
planets in the system are considerable.  In the case of fixed
Keplerian ellipses, the orbital elements $P_{1,2}$, $e_{1,2}$,
$T_{1,2}$, $\omega_{1,2}$, and $K_{1,2}$, are constant, and these,
along with the mass of the star, serve to determine completely the
positions and velocities of the planets at any time in the coplanar
case. When the interactions between the planets are non-negligible,
however, the orbital elements change continuously, and so one must
also specify an initial epoch in order to determine the motion at
future times.  Every starting epoch corresponds to a different
resultant three-body motion.

In Figure 1, we demonstrate this effect by comparing a synthetic
radial velocity curve generated by the orbital elements given in Marcy
et al. (2001) with one generated by an N-body integration using the
same elements as an initial condition. We used an epoch of JD
2450106.2, corresponding to the reported time of perihelion passage
for the outer planet.  The red line in Figure~1 shows the radial
velocity curve which results from the superposition of the two
Keplerian reflex motions. The black line shows the radial velocity
curve resulting from the full three-body integration. After three
orbits of the outer planet, the motion begins to deviate noticeably
from the dual-Keplerian approximation. After several years, the
motions have diverged completely.

The rest of this paper is organized as follows: in \S 2 we show that
self-consistent radial velocity curves are required for systems such
as GJ 876.  In \S 3, we derive improved, fully self-consistent 
dynamical fits to the observed radial velocities of the
GJ 876 system. In \S 4, we show how dual-Keplerian fits can be
improved using an
approximate analytic model for the interactions between two massive
planets in resonant systems. Further applications are discussed in \S
5, which also serves as a conclusion.

\section{Dual Keplerian vs. Self-Consistent Fits} 

The orbital elements given in Table 1 (taken from Marcy et al. 2001,
based on 54 Keck and 16 Lick observations) were derived under the
assumption that they are constants of the motion. However, for a
system such as GJ 876, where the mutual planetary interactions are
strong, the elements will change quite rapidly on observable
timescales.  We can therefore regard the parameters in Table 1 as a
set of instantaneous ``osculating'' elements, which, given a
particular starting epoch, correspond to a uniquely determined initial
condition.

Even with the assumption that $\sin i = 1$
for both planets, the orbital parameters in Table 1 lead to
a wide variety of evolutions depending on the choice of epoch.
For some starting epochs, the planets are not in the 2:1
resonance, and the system experiences severe dynamical instabilities
within five years.  For other starting epochs, the planets undergo
librations about the resonance, and the system is stable over long timescales.
One can thus ask: are there any starting epochs for which the
osculating elements in Table 1 generate an evolution which is
consistent with the observed reflex velocity of the star?

We have computed 10,000 synthetic radial velocity curves using N-body
integrations of the osculating elements for the fit to the Keck data.
Each integration used a different starting epoch spaced one day apart.
The synthetic radial velocities were given a uniform offset in
order to match the first Keck velocity point obtained by Marcy et al
(2001).  We then compared each synthetic curve to the remaining 53
Keck velocity observations, and computed a reduced ${\chi_{\nu}}^2$
statistic for the fit. The best fit occurred for an integration
with a starting epoch of JD 2450722.82. The reduced
${\chi_{\nu}}^2$ value for this fit is 12.14, and the rms scatter of
the velocities about the curve is 64 ${\rm m} {\rm s}^{-1}$.  Given
that the observational errors lie in the range 3--5 ${\rm m} {\rm
s}^{-1}$, this degree of scatter indicates that the best
dual-Keplerian fit is a poor match to the observed velocities when
mutual planetary perturbations are taken into account.

\section{A Self-consistent N-body Minimization Scheme} 

It is essential to include
mutual planetary perturbations when making fits to velocity observations
of planetary systems resembling GJ 876. One can make a
fully self-consistent fit which employs N-body integrations
to produce a synthetic reflex velocity curve for the central star.
Starting with the best dual-Keplerian fit to the combined Keck and Lick
radial velocity data sets,
we have used a Levenberg-Marquardt algorithm (Press et al 1992) to
iterate an improvement to the osculating orbital elements reported in
Table 1. Our algorithm examines how the 
${\chi_{\nu}}^2$ value of the fit depends on variations of all 10 orbital
elements, and then locates a set of elements for which the reduced
${\chi_{\nu}}^2$ fit is at a local minimum. We assume that the planets are
in co-planar orbits, and we allow the overall inclination angle of the
system to vary. 

We have found a self-consistent model for the radial velocity data
which has a reduced ${\chi_{\nu}}^2$ value of 1.46 and an rms scatter
of 13.95 ${\rm m} {\rm s}^{-1}$. This fit is shown in Figure 2 (black
curve), and represents a significant improvement to the best
dual-Keplerian fit reported by Marcy et al (2001) which had
${\chi_{\nu}}^2$=1.88.  The osculating elements are given in Table 2.
We note that the combined radial velocity data favors a co-planar
inclination of $\sin i=0.775$ for the system. There is, however, a
broad minimum around this best-fit value. At $\sin i=0.55$, the
${\chi_{\nu}}^2$=1.47, and the rms scatter is 14.69 ${\rm m} {\rm
s}^{-1}$.  Only for $\sin i<0.50$ does the fit begin to worsen
significantly.

The combined Lick and Keck data sets comprise our best overall view of
the GJ876 system. The 13.95 ${\rm m} {\rm s}^{-1}$ scatter of our best
self-consistent fit to this data is due primarily to the large 
uncertainties in the Lick velocities. It is therefore of interest to
independantly examine self-consistent fits to the Keck data only.
Applying our Levenberg-Marquardt code to the 54 Keck velocities, we
obtain a best fit with an rms scatter of 6.86 ${\rm m} {\rm s}^{-1}$,
and ${\chi_{\nu}}^2$=1.59. In this case, the minimum ${\chi_{\nu}}^2$
value occurs for an inclination $\sin i=0.55$, and the fit displays
a stronger minimum at this particular value. The orbital elements
are given in Table 2.

\section{Improved Analytic Approximations} 

The Levenberg-Marquardt minimization method
rapidly locates a dynamical fit to the radial velocity data for which
the ${\chi_{\nu}}^2$ value is at a local minimum. There is no guarantee,
however, that this minimum is global. This has motivated us to develop
an independant analytic model for fitting resonant systems. Agreement
between the two methods can thus increase our confidence that the true
configuration of the system has been located.

We start with a dual-Keplerian model using Jacobi coordinates,
where the inner planet moves on an orbit about the star and the outer
planet moves on an orbit around the center of mass of the inner two
bodies. We assume that the planets are undergoing sinusoidal
librations about the 2:1 mean motion resonance in antiphase to
each other. The semi-major axes $a$ are thus given by:
\begin{eqnarray}
a_1&=&\bar{a}_1[1+\Delta_1\cos{(n_{\rm res}t+\phi_{\rm res})}] \nonumber \\
a_2&=&\bar{a}_2[1-\Delta_2\cos{(n_{\rm res}t+\phi_{\rm res})}]
\end{eqnarray}
where the amplitude $\Delta_1$, and initial phase $\phi_{\rm res}$ are
treated as model parameters, in addition to the mean semi-major axis
$\bar{a}_2$ of the outer planet. Conservation of energy requires that
\begin{equation}
\Delta_2=\Delta_1\left(\frac{m_0m_1\bar{a}_2}
{(m_0+m_1)m_2\bar{a}_1}\right)
\end{equation}
while the resonance frequency, $n_{\rm res}$ is given approximately
by $n_{\rm res}^2\sim 3e_1n_1^2(m_2/m_0)\alpha|f_d(\alpha)|$
for the 2:1 resonance in the case where $m_2$ is significantly
larger than $m_1$ (Murray and Dermott 1999, Eqs. 8.47 and 8.32). Here
$\alpha|f_d(\alpha)|\sim 0.750$ is a constant that depends only on
the resonance involved.

The average mean motions $\bar{n}_1$ and $\bar{n}_2$ of the planets
are related by the condition that the rate of change of the resonance
critical argument is zero at exact resonance. This implies that
$\bar{n}_1-2\bar{n}_2+2(\dot{\pi}_1-\dot{\pi}_2)=0$,
where $\pi_1$ and $\pi_2$ are the longitudes of periastron of
each planet. 
The initial mean anomaly $M_2(0)$ of the outer planet is treated as a
model parameter, and the initial mean anomaly of the inner planet
$M_1(0)$ is then given by the critical argument for the resonance
\begin{equation}
\sigma=M_1-2M_2+2(\pi_1-\pi_2)
\end{equation}
where the initial value of $\sigma$ is constrained by the initial
values of $a_1$ and $a_2$.

At time $t$, the mean anomaly of body $i$ is given by
\begin{equation}
M_i=M_i(0)+\int_0^t n_i dt
\end{equation}
These integrals are straightforward to evaluate since for each 
planet $a$, and hence $n$, is an analytic function of time.

In this system, the mutual planetary perturbations are sufficiently
strong that the longitudes of periastron will precess rapidly.  We
model this by allowing each periastron longitude to vary linearly with
time, where the precession rates $\dot\pi_1$ and $\dot\pi_2$
are additional free parameters. 
In practice, we checked that the precession rates derived by the model
approximately matched those from an N-body integration using the
same initial conditions. Finally, the orbits of the planets are again assumed 
to be co-planar, but $\sin i$ of this plane is included as a parameter.

We used the analytic model to generate synthetic radial velocities for
the central star and compared these with the Keck and Lick observations
given in Marcy et al. (2001).
We initially generated a randomized population of sets of model 
parameters and then used a genetic algorithm to evolve promising sets
towards an improved description of the system. At each generation,
the genetic algorithm evaluates the degree of fit for
each parameter set, and cross breeds the best members of the population
to produce a new generation.

Figure 2 (red curve) and Table 3 report an analytic model fit for the 
combined Keck and Lick data set generated by the genetic algorithm.  
The rms scatter in this case is 13.5 ${\rm m} {\rm s}^{-1}$, with reduced
$\chi_\nu^2$ of 1.47. This represents a substantial improvement on the
best dual-Keplerian model in Table~1 (rms scatter of 21.9 ${\rm m}
{\rm s}^{-1}$).  The fit is remarkably similar to that derived
independently using the Levenberg-Marquardt algorithm.  The osculating
elements of a model fit generated by the genetic algorithm for the
Keck data alone are also given in Table 3. This fit has an rms scatter of
7.0 ${\rm m} {\rm s}^{-1}$ and a reduced $\chi_\nu^2$ of 1.67, which is
comparable to the fit obtained by the Levenberg-Marquardt N-body
technique.  As with the Levenberg-Marquardt scheme, values of $\sin i$
ranging from 0.8 to 0.5 yield solutions that are similarly good fits
to the data.

\section{Discussion}

The most important benefit of self-consistent dynamical fitting
techniques for multi-planet systems is the ability to break the $\sin
i$ degeneracy and determine the true masses of the extrasolar
planets. In a few systems, the true masses can be found when a planet
transits the parent star, but these cases will be rare and confined
largely to planets with short periods. The techniques described in \S
3 and \S 4 can in principle be applied to any system containing more
than one planet, given a sufficient baseline of observation. (For
short baselines, our techniques are most applicable to systems having
massive short-period planets.) Fischer et al 2000 have shown that
roughly half of the planetary systems found in the Lick Radial survey
show evidence of a second companion. Thus we expect that numerous
additional multi-planet systems will be forthcoming.  In the known
cases, N-body integrations of the $\upsilon$ Andromedae and HD 168443
systems indicate that planetary interactions are already producing
observable deviations from the multiple Keplerian fit.

Peale (1994) has used a roughly similar analysis to the one
described here to determine the true masses and inclinations of the
planets orbiting pulsar PSR B1257+12.  However, in that case, a
superposition of Keplerian fits provides a good approximation to
the observed reflex velocity of the pulsar. This stands in marked
contrast to cases such as GJ 876, where the planetary interactions are
an integral component of the overall motion of the star, and an
analysis based on small perturbations to these Keplerian fits may not
necessarily succeed.

Direct integration shows that 
the systems represented by the orbital parameters in Tables 2 and 
3 are stable for at least $10^7$ years, despite the presence of
strong planetary
interactions. N-body integrations of the fits to the combined Keck plus
Lick data show smooth librations about the 2:1 mean-motion resonance
and also librations about a secular resonance in which the apses
precess at the same rate (see Figure 3). These resonances help to
maintain stability by preventing close encounters between the planets.

This long-term stability, combined with the remarkable similarity
between the fits in Tables 2 and 3, derived using entirely indepedent
methods, suggests that these orbital parameters are close to
those of the real GJ 876 system. The main uncertainty remains the
precise value of $\sin i$, with our coplanar fits suggesting values 
in the range 0.5 to 0.8. The planets may also possess a mutual
inclination. We have not modeled this possibility.
Extending the baseline of observations
will make it possible to further refine the orbital fit. This, however,
may take some time. For example, Figure 4 shows the extent to which
different fits to the Keck data produced by the Levenberg-Marquardt scheme will
diverge in future. The black and red curves show the deviation of
fits with $\sin i=0.5$ and 0.7 from the best fit with $\sin i=0.55$.
The differences between these fits will remain less than the 
$3\sigma\simeq 15$ ms$^{-1}$ observational errors for several years
to come.

\subsection{Acknowledgements} 

Just prior to submitting this paper we became aware that Eugenio
Rivera and Jack Lissauer are developing a scheme similar to the
Levenberg-Marquardt procedure outlined above in order to model
the effects of mutual planetary perturbations on the Doppler
velocity variations of GJ 876.
 
We would like to thank Fred Adams, Debra Fischer, Jack Lissauer, Geoff
Marcy, Stan Peale, Eugenio Rivera and Adriane Steinacker for useful 
discussions.
This work was supported by NRC and by a NASA astrophysics theory
program which supports a joint Center for Star Formation Studies at
NASA-Ames Research Center, UC Berkeley and UC Santa Cruz.

\newpage
\begin{center}
\large Figure Captions
\end{center}

\figcaption{Synthetic radial velocity variations for GJ 876 assuming
a superposition of 2 fixed Keplerian motions for planets with elements
given in Table 1 (red line), and an N-body integration using the
same elements (black line).
\label{fig1}}

\figcaption{Synthetic radial velocity curves for the best fits
generated by the Levenberg-Marquardt scheme (black) and the
improved analytic model (red) for the combined Keck plus Lick
data. The blue circles show the observations.
\label{fig2}}

\figcaption{The evolution of the critical arguments for the 2:1
mean-motion (solid curve) and secular (dotted curve) resonances
obtained using N-body integrations of the orbital parameters
in Table 2 (black) and Table 3 (red) for the combined Keck plus
Lick data.
\label{fig3}}

\figcaption{The future divergence of fits to the Keck data by the
Levenberg-Marquardt scheme for different values of $\sin i$.
The horizontal black line shows the best fit ($\sin i=0.55$).
The black and red curves show the deviation from this fit for
fits with $\sin i=0.5$ and 0.7 respectively. The horizontal
dotted lines show typical $3\sigma=15$ ms$^{-1}$ observational errors.
\label{fig4}}

\newpage

\begin{table}
\begin{center}
\begin{tabular}{lcccc}
\hline
\hline
\\
 Parameter &  Inner & Outer & Inner & Outer \\
           & \multicolumn{2}{c}{(Keck)} & 
             \multicolumn{2}{c}{(Keck \& Lick)} \\
\\
\hline
\\
 Period $P$ (day)          & 30.11    & 61.02 & 30.12 & 61.02 \\
 $K$ (ms$^{-1}$)       & 81      & 211  & 81    & 210   \\
 Eccentricity $e$         & 0.29    & 0.11 & 0.27  & 0.10  \\
 $\omega$ (deg)        & 329     & 328  & 330   & 333   \\
 Periastron Time $T$ (JD)  & 2450031.4 & 2450105.8 & 2450031.4 & 2450106.2 \\
\\
\hline
\end{tabular}
\end{center}
\caption{\bf Best-fit dual-Keplerian elements for GJ876
(from Marcy et al. 2001)}
\end{table}

\newpage

\begin{table}
\begin{center}
\begin{tabular}{lcccc}
\hline
\hline
\\
 Parameter & Inner & Outer & Inner & Outer \\
   & \multicolumn{2}{c}{(Keck)} & \multicolumn{2}{c}{(Keck \& Lick)} \\
\\
\hline
\\
 Period\ (day)      & 29.995    & 62.092 & 30.569    & 60.128 \\
 $K$ (ms$^{-1}$)    & 85.256    & 203.16 & 84.931    & 205.351 \\
 Mass (M$_J$)       & 1.06      & 3.39   & 0.766     & 2.403  \\
 $a$\ (AU)          & 0.1294    & 0.2108 & 0.1309    & 0.2061 \\
 Eccentricity       & 0.314     & 0.051  & 0.244     & 0.039 \\
 $\omega$ (deg)     & 51.8      & 40.0   & 159.1     & 163.3 \\
 Mean anomaly (deg) & 289       & 340    & 356       & 173   \\
 Epoch (JD)         & \multicolumn{2}{c}{2450602.0931}  &
                      \multicolumn{2}{c}{2449679.6326} \\
 offset (ms$^{-1}$) & \multicolumn{2}{c}{--}            & 
                      \multicolumn{2}{c}{106.38}  \\
 $\sin i$           & \multicolumn{2}{c}{0.55}          & 
                      \multicolumn{2}{c}{0.78}  \\
\\
\hline
\end{tabular}
\end{center}
\caption{\bf Osculating orbital elements derived by Levenberg-Marquardt
N-body integration scheme.}
\end{table}

\newpage

\begin{table}
\begin{center}
\begin{tabular}{lcccc}
\hline
\hline
\\
 Parameter & Inner & Outer & Inner & Outer \\
   & \multicolumn{2}{c}{(Keck)} & \multicolumn{2}{c}{(Keck \& Lick)} \\
\\
\hline
\\
 Mass (M$_J$)       & 0.861     & 3.16   & 0.920     & 3.08  \\
 $a$\ (AU)          & 0.1290    & 0.2070 & 0.1291    & 0.2067 \\
 Eccentricity       & 0.258     & 0.054  & 0.252     & 0.046 \\
 $\omega$ (deg)     & 206       & 217    & 210       & 199   \\
 Mean anomaly (deg) & 254       & 296    & 248       & 314   \\
 Epoch (JD)         & \multicolumn{2}{c}{2450000.0}  &
                      \multicolumn{2}{c}{2450000.0} \\
 $\sin i$         & \multicolumn{2}{c}{0.60} & 
                    \multicolumn{2}{c}{0.61}  \\
\\
\hline
\end{tabular}
\end{center}
\caption{\bf Osculating astrocentric elements derived by the analytic
model plus genetic algorithm scheme.}
\end{table}

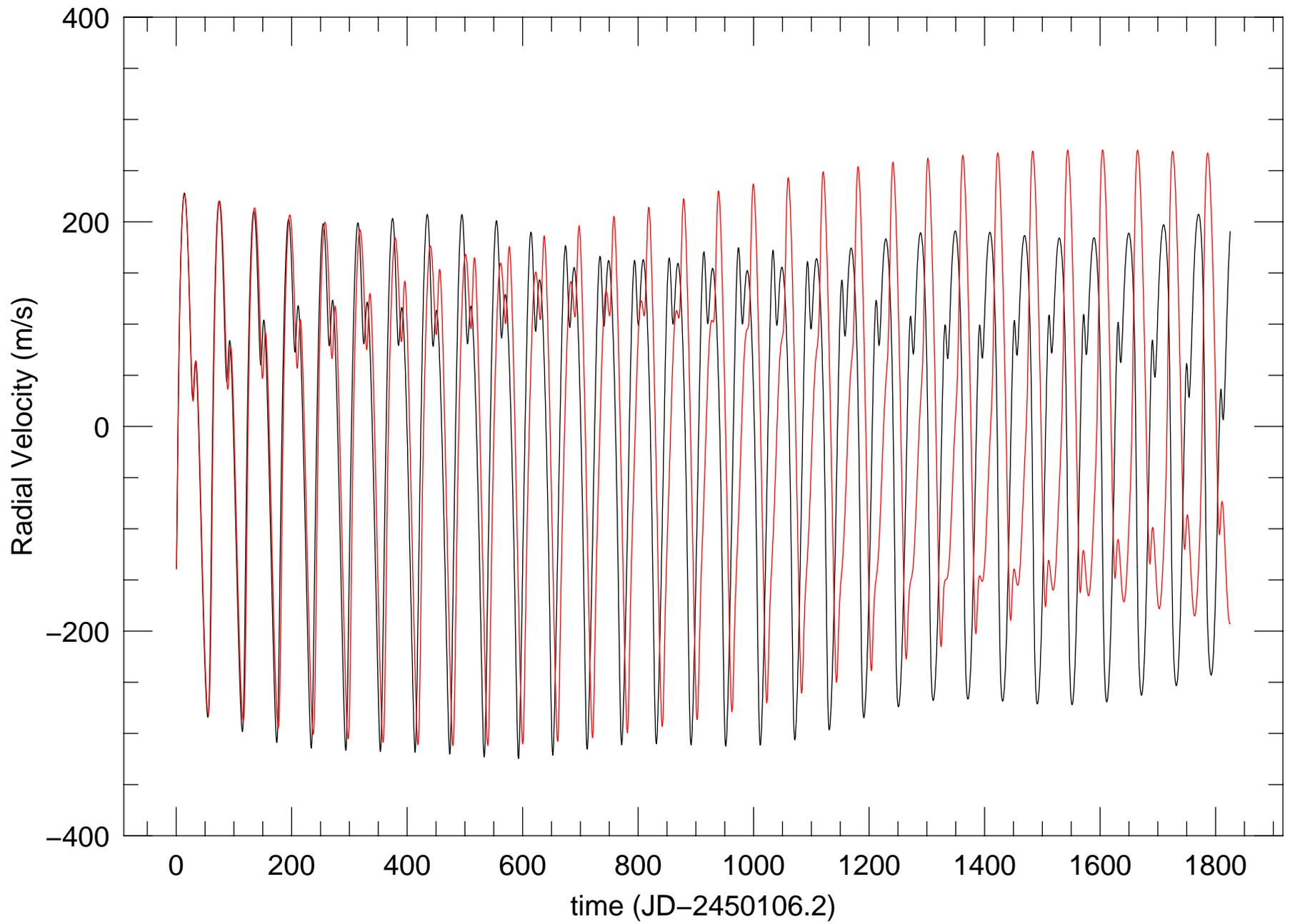
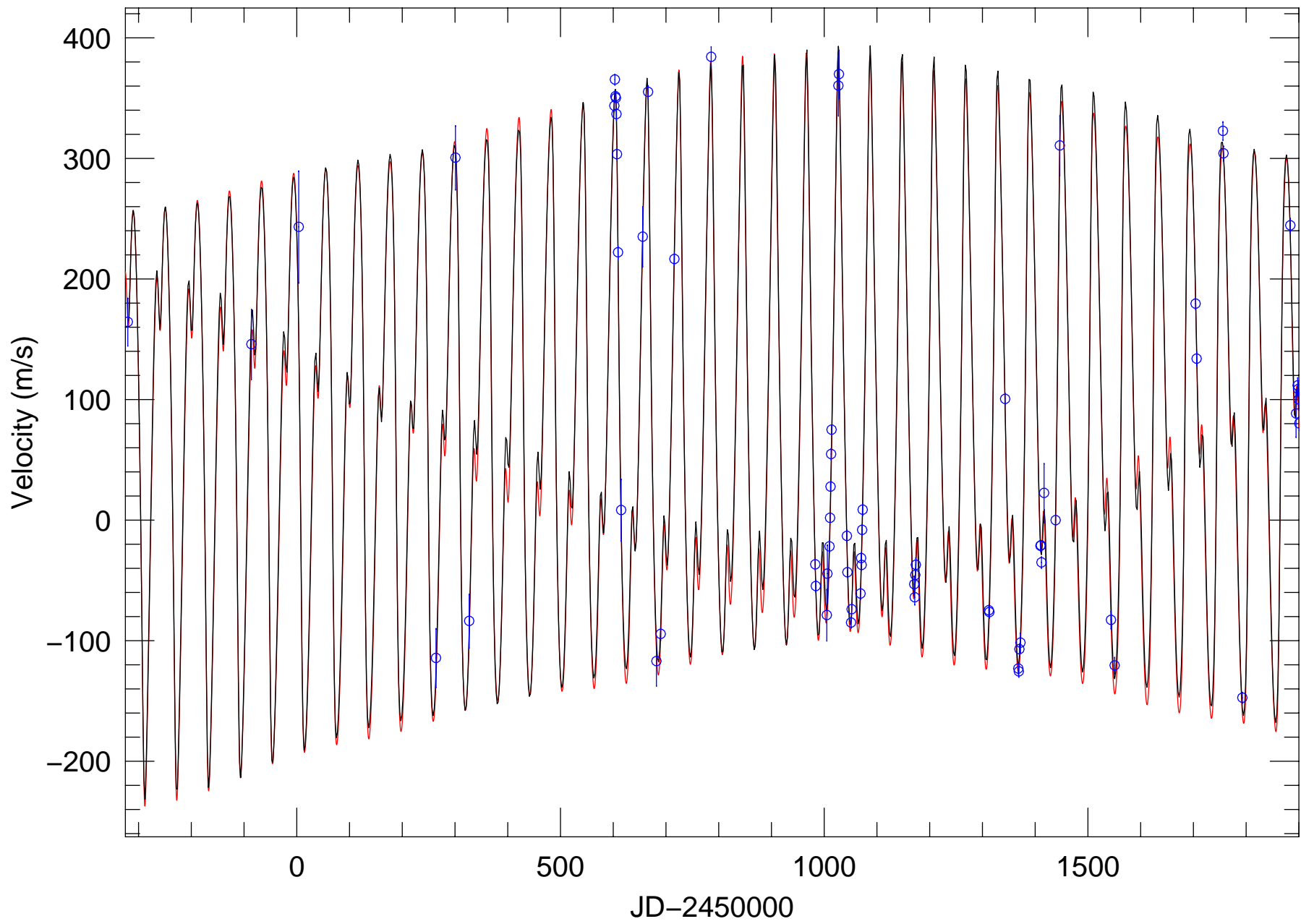
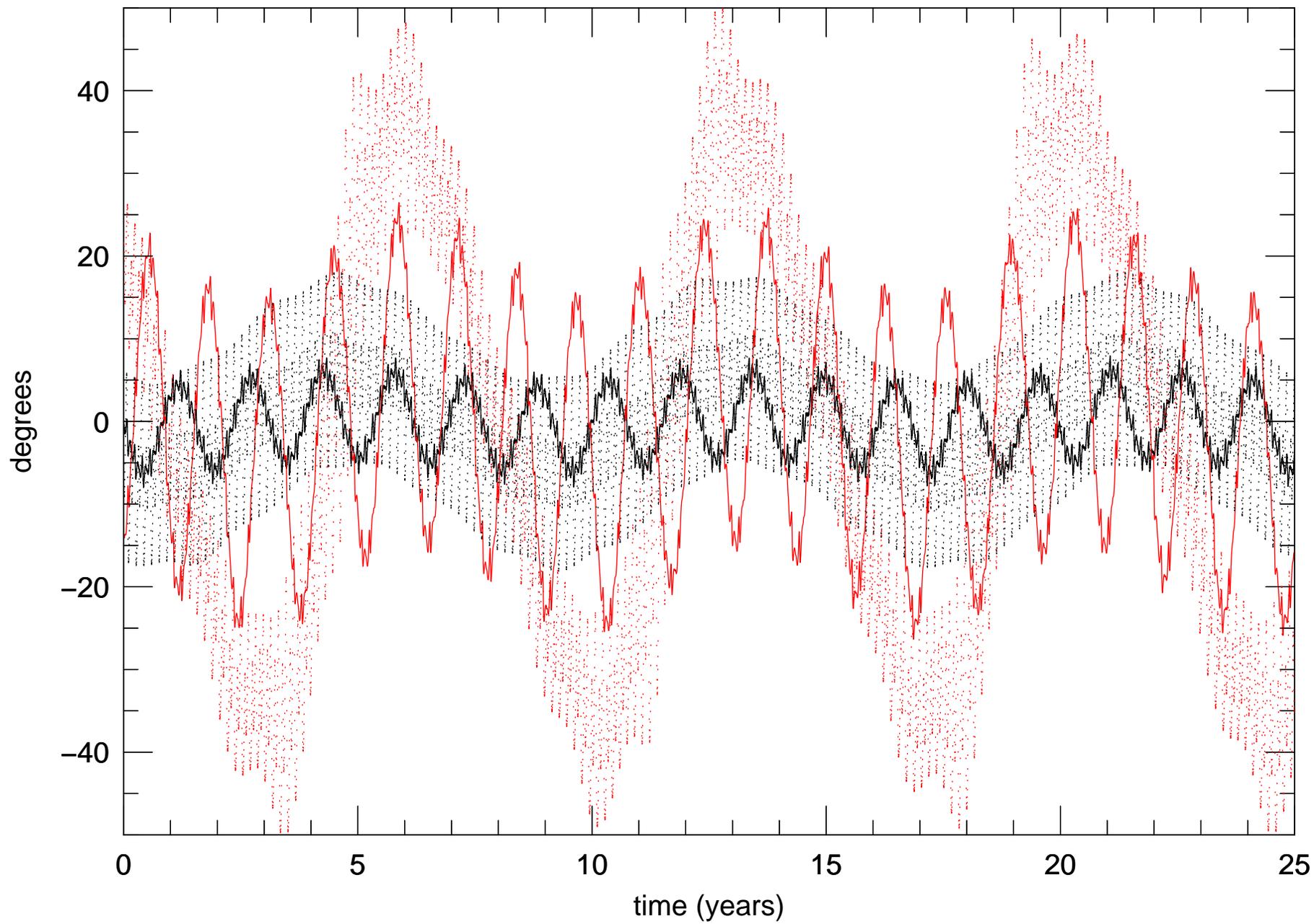
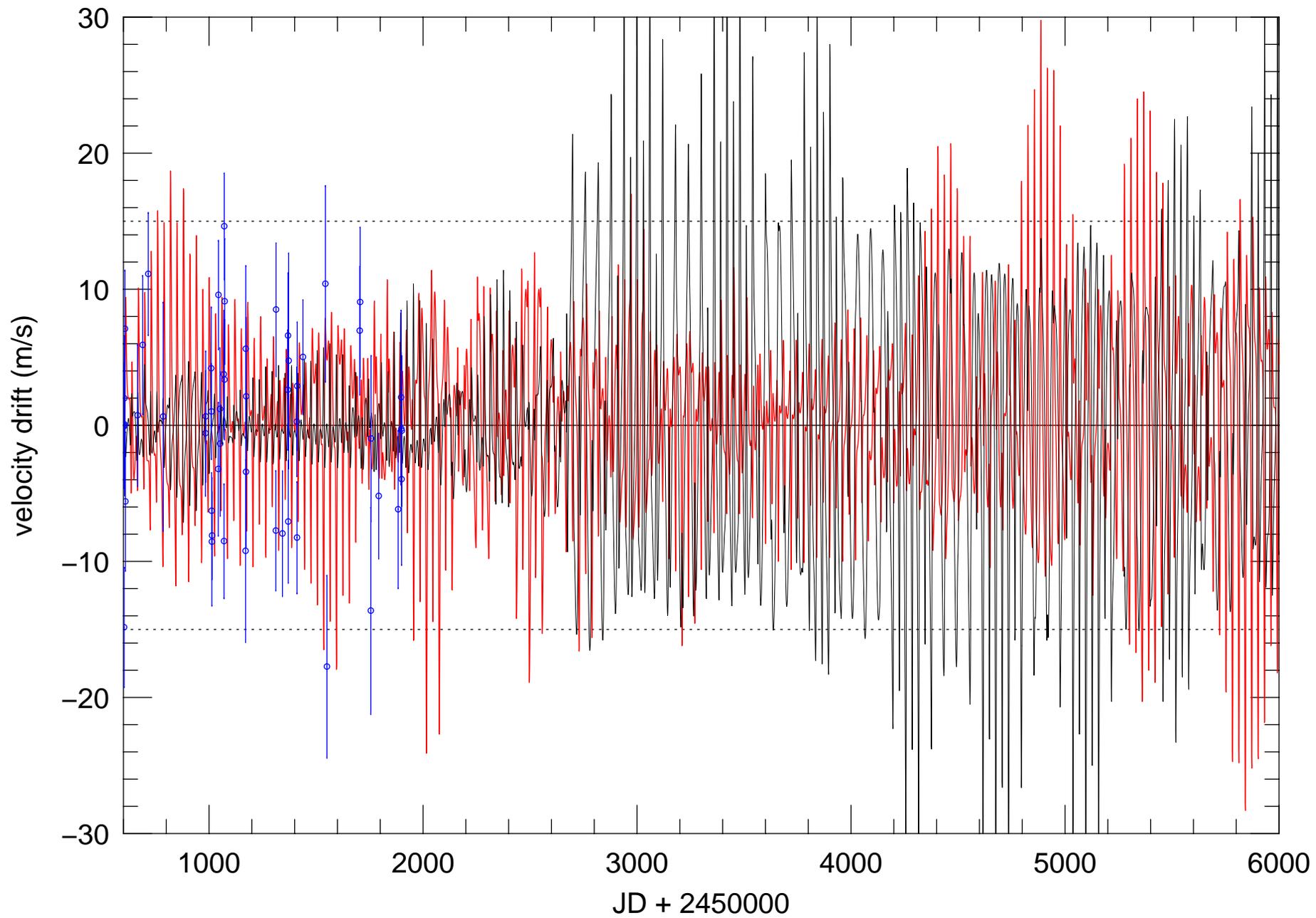
\end{document}